\documentstyle[prl,aps,preprint]{revtex}
\let\ssection=\section
\renewcommand{\section}{\setcounter{equation}{0}\ssection}
\begin{document}
\draft
%\twocolumn
%\widetext

\title{On a model of an unconstrained hyperfluid}

\author{Yuri N.~Obukhov\footnote{On leave from: 
Department of Theoretical Physics, Moscow State University, 117234 Moscow, 
Russia}}

\address{Institute for Theoretical Physics, University of Cologne,
D--50923 K\"oln, Germany}

%\date{\today}

\maketitle

\begin{abstract}
\medskip
A hyperfluid is a classical continuous medium carrying hypermomentum. We 
modify the earlier developed variational approach to a hyperfluid in such a 
way that the Frenkel type constraints imposed on the hypermomentum current 
are eliminated. The resulting self--consistent model is different from
the Weyssenhoff type one. The essential point is a conservation of the 
hypermomentum current such that the final metrical and canonical 
energy--momentum forms coincide.
\end{abstract}
\bigskip\bigskip

\section{Introduction}
\bigskip
%\bigskip

Recently \cite{1} the variational model of a hyperfluid has been developed. 
This is a classical continuous medium, elements of which possess both 
polarizability and elasticity properties. Such a matter might be a 
non-quantum source of the metric--affine gravity \cite{2} within the 
framework of the gauge theory for the general affine space-time symmetry 
group. The hyperfluid model in \cite{1} is constructed as a natural 
generalization of the phenomenological Weyssenhoff spin fluid \cite{3}, the 
variational theory of which was elaborated in \cite{4} (in the {\it 
Lagrange} approach) and in \cite{5} (in the {\it Euler} approach). Although 
the Weyssenhoff model, in our opinion, provides a reasonable description 
for a spin fluid, it is well known that there exist different models of 
continuous media with spin \cite{6,7,8}. Likewise, the properties of matter 
with hypermomentum  are not sufficiently well known at present. Thus a 
study of possible different hyperfluid models appears to be a necessary 
step in the development of this subject.  

In the present paper we discuss a possibility of removing the standard 
Frenkel type constraints usually imposed on the hypermomentum current.

\bigskip
\section{Preliminaries: matter currents in metric-affine gravity}
\bigskip

Our basic notations and conventions are that of the review \cite{2}, except 
for the metric signature which assumed here to be $(+,-,-,-)$. The field 
equations of the metric--affine gravity theory are derived from the general 
Lagrangian four-form $L = V + L_{\rm mat}$, with $V=V$($g_{\alpha\beta}$,
$\vartheta^{\alpha}$, $Q_{\alpha\beta}$, $T^{\alpha}$, 
$R_{\alpha}^{\ \beta}$) and $L_{\rm mat}=L_{\rm mat}(g_{\alpha\beta}, 
\vartheta^{\alpha}, \Gamma_\alpha{}^\beta , \Psi, D\Psi )$ as the pure 
gravitational and matter Lagrangians, respectively. Here the metric 0--form 
$g_{\alpha\beta}$, coframe 1--form $\vartheta^\alpha$, and affine connection 
1--form $\Gamma_\alpha{}^\beta$ are the gravitational potentials, the 
respective field strengths are nonmetricity 1-form $Q_{\alpha\beta}:=- 
Dg_{\alpha\beta}$ and the 2-forms of torsion $T^\alpha$ and curvature 
$R_\alpha{}^\beta$ \cite{2}. The general field equations read
\begin{equation}
2{\delta V \over \delta g_{\alpha\beta}}=-\sigma^{\alpha\beta},\quad
{\delta V \over \delta \vartheta^{\alpha}}=-\Sigma_{\alpha},\quad
{\delta V \over \delta \Gamma_{\alpha}^{\ \beta}}=-
\Delta^{\alpha}_{\ \beta},\label{2.1}
\end{equation}
where the matter currents are defined as the variational derivatives
\begin{equation}
\sigma^{\alpha\beta}:= 2{{\delta L_{\rm mat}}\over{\delta g_{\alpha\beta}}},
\quad\Sigma_{\alpha}:={{\delta L_{\rm mat}}\over{\delta\vartheta^{\alpha}}},
\quad\Delta^{\alpha}{}_{\beta}:= 
{{\delta L_{\rm mat}}\over{\delta\Gamma_{\alpha}{}^{\beta}}}.\label{2.2}
\end{equation}
It worthwhile to mention that the standard Einstein-Hilbert Lagrangian $V=-
{1\over 2}R_{\alpha}{}^{\beta}\wedge\ast(\vartheta^{\alpha}\wedge
\vartheta_{\beta})$ (considered in the recent paper \cite{9}, e.g.) is a 
bad choice for metric--affine gravity. Due to an auxilliary (so-called 
projective, see \cite{2}) invariance, this Lagrangian is only compatible 
with the dilaton-free sources for which $\Delta =
\Delta^{\alpha}{}_{\alpha}=0$, and the Weyl 1-form 
$Q=g^{\alpha\beta}Q_{\alpha\beta}$ is not determined by the 
gravitational field equations (\ref{2.1}).

The Weyssenhoff type hyperfluid \cite{1} is described by the following 
currents:

\noindent the metric stress-energy 4-form
\begin{equation}
\sigma^{\alpha\beta}=\eta\big((\varepsilon + p) u^{\alpha}u^{\beta} - 
pg^{\alpha\beta}\big)+ 2u^{\gamma}u^{(\alpha}D\Delta^{\beta)}_{\ \gamma},
\label{2.3}
\end{equation}
the canonical energy-momentum 3-form
\begin{equation}
\Sigma_{\alpha}=uP_{\alpha}- p(\eta_{\alpha}-uu_{\alpha}),\label{2.4}
\end{equation}
and the hypermomentum 3-form
\begin{equation}
\Delta^{\alpha}_{\ \beta}=uJ^{\alpha}_{\ \beta},\label{2.5}
\end{equation}
where $\varepsilon$ is the energy density, $p$ the pressure, $u$ the
flow $3$-form, $J^{\alpha}_{\ \beta}$ the {\it hypermomentum density}, 
and $P_\alpha$ the four-momentum 
\begin{equation}
P_{\alpha}= \ast (\varepsilon u\wedge\vartheta_{\alpha} - 
2u^{\beta}g_{\gamma [\alpha }D\Delta^\gamma {}_{\beta ]}).\label{2.6}
\end{equation}
As usually, $\eta^{\alpha}=\ast\vartheta^{\alpha}$ and $\eta$ is the volume 
$4$-form. Components of $4$-velocity are defined by $u =
u^{\alpha}\eta_{\alpha}$. Hypermomentum includes as irreducible 
parts the spin density $S_{\alpha\beta}=J_{[\alpha\beta]}$ and the dilaton
charge $J=J^{\alpha}{}_{\alpha}$.

Hypermomentum density satisfies the constraint,
\begin{equation}
J^{\alpha}_{\ \beta}u^{\beta}=J^{\alpha}_{\ \beta}u_{\alpha}=0.\label{2.7}
\end{equation}
This is a natural generalization of the Frenkel condition $S_{\alpha\beta}
u^{\beta}=0$.  

Recently \cite{10} it has been suggested that the generalised Frenkel 
condition could be weakened and, in particular, reduced just to the 
ordinary Frenkel constraint imposed on the spin part of the hypermomentum 
density. Below we study the possibility of constructing the completely 
{\it unconstrained} hyperfluid model. 

\bigskip
\section{Variational principle for the unconstrained hyperfluid}
\bigskip

We start from the Lagrangian 4-form for the hyperfluid, cf. \cite{1},
$$
L_{\rm mat}=\varepsilon(\rho, s, \mu^{A}_{\ B})\eta - 
{1\over 2}\rho\mu^{A}_{\ B}b^{B}_{\alpha}u\wedge Db^{\alpha}_{A} -
$$
\begin{equation}
- \rho u\wedge d\lambda_{1} + \lambda_{2}u\wedge dX +
\lambda_{3}u\wedge ds + \lambda_{0}(\ast u\wedge u - \eta) 
+ \lambda^{A}_{B}(b^{B}\wedge b_{A} - \eta\delta^{B}_{A}).\label{3.1}
\end{equation}
Here the first two terms describe the internal energy density 
$\varepsilon$ and the kinetic energy density, respectively. As usually, 
we assume that the internal energy of hyperfluid depends on the particle 
density $\rho$, the specific entropy $s$ and the specific hypermomentum 
density $\mu^{A}{}_{B}$. It seems worthwhile to notice at this point that 
the indices $A,B,...$, which run from 1 to 3, are merely labels and they 
have no any geometrical meaning. The material frames $b^A$ (which is a 
1-form with an expansion $b^{A}=b^{A}_{\alpha}\vartheta^\alpha$) and $b_A$ 
(which is a 3-form with analogous expansion $b_{A}=b_{A}^{\alpha}
\eta_\alpha$) have nothing to do with the {\it reference} frames which an 
observer may introduce and change (i.e. rotate and deform) at his own 
choice. The material frames together with the specific hypermomentum are 
a sort of internal variables which describe the polarizability and 
elasticity properties of the fluid elements. Material frames are rigidly 
attached to elements of the continuum and their evolution is determined 
by the equations of motion of the fluid. From the geometrical point of 
view, $\mu^{A}{}_{B}$ is a scalar and $\varepsilon$ is explicitly a 
generally coordinate invariant quantity. The energy density cannot be a 
function of the hypermomentum density {\it tensor}, as is incorrectly 
assumed in \cite{9}, since this is incompatible with the general 
coordinate invariance of the fluid action. 

The remaining terms in (\ref{3.1}) describe constraints. The first three 
express the conservation particle number, constancy of entropy, and 
identity of elements along the flow lines:
\begin{equation}
d(\rho u)=0,\label{3.2}
\end{equation}
\begin{equation}
u\wedge dX=0,\label{3.3}
\end{equation}
\begin{equation}
u\wedge ds=0.\label{3.4}
\end{equation}

The last two terms in (\ref{3.1}) tells that the fluid velocity is 
timelike and normalized,
\begin{equation}
\ast u\wedge u=\eta,\label{3.5}
\end{equation}
and that material frame variables are dual in the sense 
\begin{equation}
b^{B}\wedge b_{A} = \delta^{B}_{A}\eta .\label{3.6}
\end{equation}
Here is the crucial modification of the hyperfluid model \cite{1}. Recall 
that the material triad is {\it rigid} in the spin fluid model (which can 
only rotate), while in accordance with the affine gauge approach we assumed
that in a hyperfluid the material frame $b^{\alpha}_{A}$ is {\it elastic} 
and can deform arbitrarily during the motion of the medium. However, in the 
Weyssenhoff type model \cite{1} the generalized material tetrad 
$(u, b_{A})$ is still constrained in that the material triad is assumed to 
be orthogonal to the velocity, $\ast u\wedge b_{A}=0$. Now, developing the 
suggestion of \cite{10}, we remove the latter orthogonality condition. 
Hereafter such a medium is called {\it unconstrained hyperfluid}.  

The derivation of the Euler-Lagrange equations is analogous to \cite{1}. 
As before, the independent variables are: the metric-affine gravitational 
potentials $g_{\alpha\beta}$,\ $\vartheta^{\alpha}$,\ 
$\Gamma^{\ \alpha}_{\beta}$, the material variables 
$\Psi =\{b_{A},\ b^{B},\ \rho, \mu^{A}_{\ B}$,\ $s,\ X\}$,
and the Lagrange multiplier 0-forms $\lambda_0, \lambda_1, \lambda_2, 
\lambda_3, \lambda^{A}_{B}$.

Varying the action (\ref{3.1}) with respect to the Lagrange multipliers, 
one finds (\ref{3.2})-(\ref{3.6}). Variations of $s, X, \rho, \mu^{A}_{\ B}$ 
and $u$, $b^{B}$, $b_{A}$ yield the equations, respectively, 
\begin{equation}
\eta\Big({\partial\varepsilon \over\partial s}\Big)+d(\lambda_{3}u)=0,\quad  
d(\lambda_{2}u)=0,\label{3.7}
\end{equation}
\begin{equation}
\eta\Big({\partial\varepsilon \over \partial\rho}\Big) -
{1\over 2}\mu^{A}_{\ B}b^{B}_{\alpha}u\wedge Db^{\alpha}_{A} -
u\wedge d\lambda_{1} = 0,\label{3.8}
\end{equation}
\begin{equation}
\eta\Big({\partial\varepsilon \over \partial\mu^{A}_{\ B}}\Big) =
{1\over 2}\rho b^{B}_{\alpha}u\wedge Db^{\alpha}_{A},\label{3.9}
\end{equation}
\begin{equation}
-{1\over 2}\rho\mu^{A}_{\ B}b^{B}_{\alpha}Db^{\alpha}_{A}-\rho d\lambda_{1} 
+ \lambda_{2}dX + \lambda_{3}ds - 2\lambda_{0}\ast\!\! u=0,\label{3.10}
\end{equation}
\begin{equation}
{1\over 2}\rho\mu^{A}_{\ B}\ast\!\!(u\wedge D b^{\alpha}_{A})
\eta_{\alpha} + \lambda^{A}_{B}b_{A}=0,\label{3.11}
\end{equation}
\begin{equation}
{1\over 2}\ast\!\! D(\rho\mu^{A}_{\ B}b_{\alpha}^{B}u)\
vartheta^{\alpha} + \lambda^{A}_{B}b^{B}=0.\label{3.12}
\end{equation}
These equations determine the Lagrange multipliers. While $\lambda_1, 
\lambda_2, \lambda_3$ satisfy differential evolution equations, one can
explicitly solve (3.8)-(3.11) with respect to the other Lagrange 
multipliers,
\begin{equation}
2\lambda_{0}=
\rho\Big({\partial\varepsilon \over \partial\rho}\Big),\label{3.13}
\end{equation}
\begin{equation}
\lambda^{A}_{B}=
{1\over 2}\rho\mu^{C}_{\ B}\ast\!\! 
(u\wedge b_{C}^{\alpha}Db_{\alpha}^{A}).\label{3.14}
\end{equation}

Substituting (\ref{3.14}) back into (\ref{3.11}) and (\ref{3.12}), we get 
the system of {\it equations of motion} of the specific hypermomentum 
density and the material frames:
\begin{equation}
u\wedge (d\mu^{A}_{\ B} + \mu^{A}_{\ C}b^{\alpha}_{B}Db_{\alpha}^{C} - 
\mu^{C}_{\ B}b^{\alpha}_{C}Db_{\alpha}^{A}) = 0,\label{3.15}
\end{equation}
\begin{equation}
(\delta^{\beta}_{\alpha} - b^{C}_{\alpha}b^{\beta}_{C})\mu^{A}{}_{B}u\wedge 
Db^{B}_{\beta} =0,\label{3.16}
\end{equation}
\begin{equation}
(\delta^{\beta}_{\alpha} - b^{C}_{\alpha}b^{\beta}_{C})\mu^{A}{}_{B}u\wedge 
Db_{A}^{\alpha} =0.\label{3.17}
\end{equation}
While (\ref{3.15}) coincides with the analogous equation of motion for the
Weyssenhoff type hyperfluid \cite{1}, equations (\ref{3.16})-(\ref{3.17}) 
are new. One can  rewrite (\ref{3.15})-(\ref{3.17}) in terms of the tensor 
variables, demonstrating that these equations express the conservation of 
hypermomentum current. Indeed, contracting (\ref{3.15}) with 
$b^{\alpha}_{A}b^{B}_{\beta}$ and using (\ref{3.16})-(\ref{3.17}),
we find
\begin{equation}
u\wedge D(\mu^{A}{}_{B}b_{A}^{\alpha}b^{B}_{\beta})=0.\label{3.18}
\end{equation}

The {\it matter currents} of the unconstrained hyperfluid are derived by 
direct calculation of variational derivatives (\ref{2.2}). Denoting as 
usually \cite{1},
\begin{equation}
J^{\alpha}_{\ \beta}= 
{1\over 2}\rho\mu^{A}_{\ B}b^{\alpha}_{A}b^{B}_{\beta},\label{3.19}
\end{equation}
and introducing the {\it pressure} in a standard way,
\begin{equation}
p:=\rho\Big({\partial\varepsilon\over \partial\rho}\Big) - 
\varepsilon ,\label{3.20}
\end{equation}
one obtains (using (\ref{3.15})-(\ref{3.18})):
\begin{equation}
\sigma^{\alpha\beta}=\eta\big((\varepsilon + p)u^{\alpha}u^{\beta} - 
pg^{\alpha\beta}\big),\label{3.21}
\end{equation}
\begin{equation}
\Sigma_{\alpha}= \varepsilon uu_{\alpha} - 
p(\eta_{\alpha}-uu_{\alpha}),\label{3.22}
\end{equation} 
\begin{equation}
\Delta^{\alpha}_{\ \beta}=u J^{\alpha}_{\ \beta}.\label{3.23}
\end{equation}

The most essential difference of the resulting unconstrained model from 
the Weyssenhoff type hyperfluid \cite{1} is the complete decoupling
of hypermomentum from both energy 3-forms (\ref{3.21}) and (\ref{3.22}). 
This is compatible with the Noether identities \cite{2} though, as the 
equation (\ref{3.18}) (multiply by $\rho$ and use (\ref{3.4})) actually 
describes the conservation of the hypermomentum current,
\begin{equation}
D\Delta^{\alpha}_{\ \beta} = 0.\label{3.24}
\end{equation}
Dynamics of all irreducible parts of hypermomentum is contained in 
(\ref{3.24}), and it is indeed unconstrained since neither the Frenkel 
type conditions (\ref{2.7}), nor any other are imposed on the 
hypermomentum density.

The complete decoupling of $\Delta^{\alpha}_{\ \beta}$ from the energy
currents of course does not mean that unconstrained hypermomentum is not
affecting the gravitational field: it still enters as the source in
the third metric-affine gravity field equation (\ref{2.1}).

At the first sight, it may seem surprising that such a simple modification
of the original hyperfluid model, as the elimination of orthogonality 
condition of the material triad and velocity, may yield the above described
essential change of the equations of motion of hypermomentum and reduce
the energy-momentum 3-form to that of the ideal structureless fluid. 
However, everything becomes clear when one notices that the Lagrangian 
(\ref{3.1}) of the unconstrained hyperfluid possesses an extra gauge 
symmetry besides the usual coordinate and local gravitational frame 
$GL(4,R)$ invariances. Namely, (\ref{3.1}) is invariant under the 
simultaneous transformations of the material frames and the affine 
connection components,
\begin{equation}
b^{\alpha}_{A}\longrightarrow L^{\alpha}_{\beta}b^{\beta}_{A},\quad
b_{\alpha}^{B}\longrightarrow b_{\beta}^{B}L^{-1\beta}_{\ \ \ \alpha},
\quad\Gamma_{\alpha}{}^{\beta}\longrightarrow L^{\beta}_{\sigma}
(\Gamma_{\rho}{}^{\sigma}+\delta^{\sigma}_{\rho}d) 
L^{-1\rho}_{\ \ \ \alpha},\label{3.25} 
\end{equation}
where $L^{\alpha}_{\beta}\in GL(4,R)$. Notice that this is different from
the affine gauge gravity transformation, since the gravitational frame 
$\vartheta^\alpha$ and metric $g_{\alpha\beta}$ are not transformed. The 
total Lagrangian $L=V+L_{\rm mat}$ certainly does not possess the symmetry
(\ref{3.25}). Applying to the Lagrangian (\ref{3.1}) the Noether machinery 
\cite{2}, one can derive a conservation law corresponding to (\ref{3.25}). 
This is exactly (\ref{3.24}).

\bigskip
\section{Discussion and conclusion}
\bigskip
  
In the Weyssenhoff type hyperfluid \cite{1} the generalized Frenkel 
condition (\ref{2.7}) plays an important role. It indeed restricts the 
possible motions of the medium, as well as the very structure of the 
hypermomentum current. In particular, (\ref{2.7}) rules out the case of 
purely dilatonic Weyssenhoff matter. Using the standard decomposition 
\cite{2} of hypermomentum into its irreducible parts, 
$J^{\alpha}{}_{\beta}={\nearrow\!\!\!\!\!\!\!J}{\ }^{\alpha}{}_{\beta}+ 
{1\over 4}J\delta^{\alpha}_{\beta} + S^{\alpha}{}_{\beta}$, one finds in 
view of (\ref{2.7}) that the dilaton charge is expressed in terms of the 
proper hypermomentum (shear) according to 
$J=-4{\nearrow\!\!\!\!\!\!\!J}_{\alpha\beta}u^{\alpha}u^{\beta}$. Thus 
vanishing shear yields also $J=0$. Unlike this, the unconstrained 
hyperfluid may be of the purely dilatonic type. In this case our model 
gives a description of a physical source for the generalized
Einstein-Weyl gravity theory considered recently in \cite{11}.

In \cite{10} an attempt was made to construct a sort of intermediate
model in which the constraint (\ref{2.7}) reduces to the original Frenkel 
condition, $F_{\alpha}=S_{\alpha\beta}u^{\beta}=0$. Technically this can 
be achieved if one adds to (\ref{3.1}) the term $\xi^{\alpha}F_{\alpha}$ 
with a Lagrange multiplier (4--form) $\xi^{\alpha}$. Such a term breaks the 
symmetry (\ref{3.25}). However the resulting equations of motion for the 
fluid and the gravitational currents (\ref{2.2}) look unusual and their 
physical interpretation is unclear. Contrary to the expectations of the 
authors of \cite{10} (who even failed to solve the highly nontrivial 
system of constraint equations with respect to $\xi^{\alpha}$), at
the end one does not find a Weyssenhoff type dynamics for the spin part of
the hypermomentum. 

It should be noted that the described above unconstrained hyperfluid is 
not a subcase of the general Weyssenhoff hyperfluid model. They are close
though, in the sense that the two theories may admit the same particular 
solutions for the gravitational and matter field equations. The next 
decisive step will be a comparison of different ideal hyperfluid models 
\cite{1,9,10} with the real physical media the elements of which display 
polarizability and/or elasticity properties. 

\bigskip
{\bf Acknowledgements}
\bigskip

I would like to thank Friedrich W. Hehl for stimulating discussions and 
useful comments. This research was supported by the Deutsche 
Forschungsgemeinschaft under the project He-528/17-1. 
%\vfill\eject

%\vfill\bye
\end{document}